\documentclass[
reprint,
superscriptaddress,
amsmath,amssymb,
prb
]{revtex4-2}

\usepackage{graphicx}
\usepackage{dcolumn}
\usepackage{bm}
\usepackage{chemformula}
\usepackage{hyperref}
\usepackage{amsfonts}
\usepackage[english]{datetime2}
\usepackage{tabularx}
\usepackage{float}
\usepackage[percent]{overpic}
\usepackage{tikz}

\usepackage{array}
\usepackage{booktabs}
\usepackage{makecell}

\usepackage[version=4]{mhchem}
\usepackage{physics}

\begin{document}
	
	\title{Crystal-Field Symmetry Constraints in Layered Honeycomb ErBr$_3$}
	
	\author{Biaoyan~Hu}
	\email{hubiaoyan@quantumsc.cn}
	\affiliation{Quantum Science Center of Guangdong-Hong~Kong-Macao Greater Bay Area, Shenzhen 518045, China}
	\affiliation{State Key Laboratory of Quantum Functional Materials $\&$ Department of Physics, Southern University of Science and Technology, Shenzhen 518055, China}
	\author{Mingyuan~Hu}
	\affiliation{State Key Laboratory of Quantum Functional Materials $\&$ Department of Physics, Southern University of Science and Technology, Shenzhen 518055, China}
	\affiliation{Guangdong Basic Research Center of Excellence for Quantum Science $\&$ Guangdong Provincial Key Laboratory of Advanced Thermoelectric Materials and Device Physics, Southern University of Science and Technology, Shenzhen 518055, China}
	\author{Franz~Demmel}
	\affiliation{ISIS Facility, Rutherford Appleton Laboratory, Didcot OX11 0QX, UK}
	\author{Andrey~A.~Podlesnyak}
	\affiliation{Neutron Scattering Division, Oak Ridge National Laboratory, Oak Ridge, Tennessee 37831, USA}
	\author{Jiaqing~He}
	\affiliation{State Key Laboratory of Quantum Functional Materials $\&$ Department of Physics, Southern University of Science and Technology, Shenzhen 518055, China}
	\affiliation{Guangdong Basic Research Center of Excellence for Quantum Science $\&$ Guangdong Provincial Key Laboratory of Advanced Thermoelectric Materials and Device Physics, Southern University of Science and Technology, Shenzhen 518055, China}
	\author{Liusuo~Wu}
	\affiliation{State Key Laboratory of Quantum Functional Materials $\&$ Department of Physics, Southern University of Science and Technology, Shenzhen 518055, China}
	\affiliation{Quantum Science Center of Guangdong-Hong~Kong-Macao Greater Bay Area, Shenzhen 518045, China}
	
	\begin{abstract}
		Crystal-field symmetry restricts the ground-state Kramers doublet of ErBr$_3$ to one of two classes. We show that the compressed octahedral environment selects the class with $\langle \psi_\pm | J^{\pm} | \psi_\mp \rangle = 0$, suppressing the lowest-order $J^{\pm}$-mediated exchange. Thermodynamic measurements reveal two zero-field anomalies at 0.375 and 0.200~K. Under an in-plane magnetic field, the thermodynamic response separates into a phase boundary and a broader crossover line. Inelastic neutron scattering measurements at 2 K reveal no well-defined low-energy dispersive magnetic modes. These results connect the ground-state symmetry with the field-dependent thermodynamic response of ErBr$_3$, providing a microscopic starting point for understanding its low-energy magnetic behavior.
	\end{abstract}
	
	\maketitle
	
	\section{Introduction}
	
	Strong spin--orbit coupling entangles spin and orbital degrees of freedom \cite{WitczakKrempa.2014,Rau.2016}. In frustrated magnets, the resulting magnetic behavior can be strongly shaped by local symmetry and competing interactions \cite{Balents.2010}. In rare-earth magnets, the crystalline electric field (CEF) splits the $J$ multiplet into a sequence of Kramers doublets \cite{Abragam.1970,hutchings1964point}. The low-temperature physics can often be described in terms of an effective pseudospin-$1/2$ degree of freedom with an anisotropic Zeeman response \cite{Chibotaru2008PRL}. In the present context, however, the microscopic magnetic response is governed by matrix elements of the total angular momentum operator $\mathbf{J}$ rather than the bare spin $\mathbf{S}$. The low-temperature properties are determined by the representation and matrix-element structure of the ground-state doublet. In particular, symmetry may force the off-diagonal matrix element of $J^\pm$ within the ground-state doublet to vanish. In a lattice magnet, this suppresses the corresponding exchange processes mediated by $J^{\pm}$ operators and constrains how low-energy magnetic fluctuations propagate from site to site, with important consequences for collective magnetic behavior.
	
	Frustrated magnets can host quantum-spin-liquid behavior and unconventional collective excitations \cite{Ross.2011,Savary.2017}. In spin--orbit-coupled honeycomb systems, Kitaev, Heisenberg, and longer-range interactions can compete to produce rich phase diagrams \cite{Winter.2016,Kimchi.2011,Rau.2016}; neutron scattering on the proximate Kitaev material $\alpha$-RuCl$_3$ provides a prominent experimental example of unusual magnetic excitations \cite{Banerjee2017}.
	
	Rare-earth honeycomb magnets have emerged as promising platforms for exploring spin--orbit-driven magnetism, because crystal-field-selected Kramers-doublet ground states provide low-energy degrees of freedom on which local symmetry and exchange interactions act \cite{Pistawala2024,Khatua2023PRB,Liu2023InorgChem}. An effective pseudospin-$1/2$ description based on a crystal-field-selected doublet is also used for spin--orbit-driven rare-earth magnets in other lattice geometries \cite{Khatua2024}. Layered rare-earth halides are particularly appealing in this context, since they realize quasi-two-dimensional honeycomb lattices and can host strongly anisotropic magnetic responses \cite{Kramer1999PRB,Kramer2000PB,Pistawala2026,Jang2024CommunMater}. Previous studies have revealed diverse low-temperature behaviors, including noncollinear ordering in Er$X_3$ compounds \cite{Kramer1999PRB,Kramer2000PB}, anisotropic magnetic ground states in YbI$_3$ \cite{Pistawala2024}, and strongly anisotropic rare-earth ground states in several other systems \cite{Zhao2023TbScO3,Tian2021DyOCl}. In ErBr$_3$, inelastic neutron scattering within the ordered phase has revealed dispersive magnetic modes \cite{Wessler2022}. However, the connection between the symmetry structure of the local CEF ground state and the field-dependent thermodynamic response has not yet been systematically established.
	
	\begin{figure*}[t]
		\centering
		\includegraphics[width=2\columnwidth]{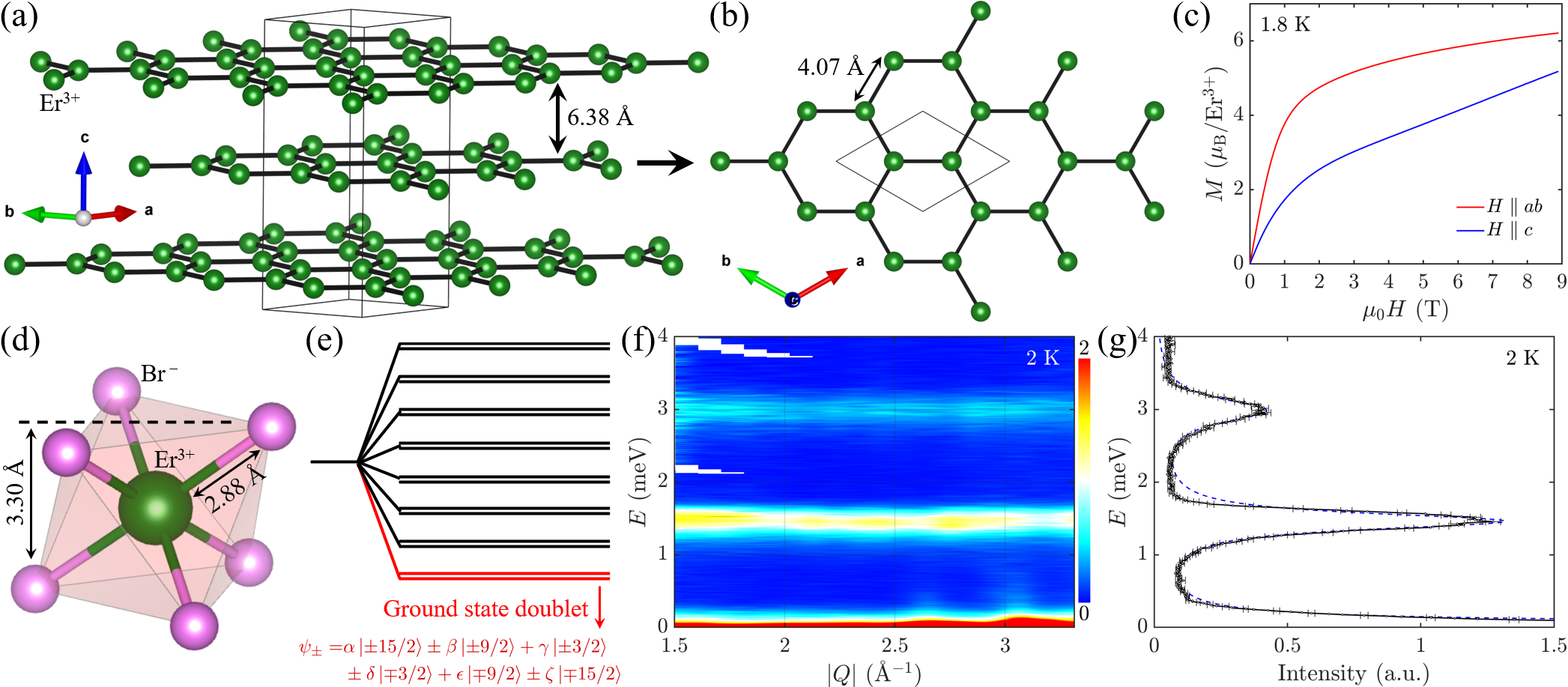}
		\caption{
			(a) Three-dimensional crystal structure of ErBr$_3$, showing stacked honeycomb layers of Er$^{3+}$ ions separated along the crystallographic $c$ axis. The interlayer spacing is indicated.
			(b) Top view of a single honeycomb layer in the $ab$ plane, with the nearest-neighbor Er--Er distance labeled.
			(c) Field-dependent magnetization measured at 1.8 K for magnetic fields applied parallel to the honeycomb plane ($H \parallel ab$) and along the crystallographic $c$ axis ($H \parallel c$), demonstrating pronounced magnetic anisotropy.
			(d) Local coordination environment of an Er$^{3+}$ ion, illustrating the ErBr$_6$ octahedron, weakly compressed along the trigonal axis. The mean Er--Br bond length and the triangular-face separation are indicated.
			(e) Schematic CEF energy-level scheme of the Er$^{3+}$ ion. The $J=15/2$ multiplet is split into a series of Kramers doublets, with the ground-state doublet highlighted in red. Its $|m_J\rangle$ composition, shown schematically below following Eq.~\ref{eq:psi0}, is given quantitatively in Table~\ref{tab:cef_levels}.
			(f) Inelastic neutron scattering intensity map measured at 2~K as a function of energy transfer and momentum transfer $|Q|$.
			(g) $Q$-integrated INS intensity measured at 2~K. Error bars, shown every third point for clarity, represent statistical uncertainties. The dashed line is a three-peak Lorentzian fit to the spectrum, resolving two dominant finite-energy crystal-field excitations at 1.46 and 2.98~meV, with no additional low-energy magnetic mode resolved.
		}
		\label{fig:structure_cef}
	\end{figure*}
	
	Here we investigate ErBr$_3$ by combining crystal-field analysis with magnetization, thermodynamic, magnetocaloric, and INS measurements. Our results relate the ground-state Kramers-doublet symmetry to the field-dependent thermodynamic response.
	
	\section{Experimental Methods}
	
	Single crystals of ErBr$_3$ were grown by the Bridgman method. We obtained a high-quality single-domain crystal with a mass of about 13~g. The crystallographic orientation was checked by x-ray Laue diffraction, and the field directions used in the thermodynamic measurements were defined with respect to the honeycomb plane ($H \parallel ab$) and the crystallographic $c$ axis ($H \parallel c$).
	
	The magnetization data shown in this work were measured on single-crystalline ErBr$_3$ at 1.8~K in fields up to 8.9~T applied both within the honeycomb plane and along the crystallographic $c$ axis. Temperature-dependent magnetization measurements were also performed to obtain a Curie--Weiss estimate. The specific-heat data discussed in this work comprise temperature sweeps on oriented single-crystalline samples shown between 0.16 and 2.4~K at selected fields up to 1~T, together with field sweeps shown between 0 and 1~T at selected temperatures between 0.1 and 1~K. Magnetocaloric-effect measurements were performed with $H\parallel ab$ during continuous decreasing-field sweeps.
	
	Inelastic neutron scattering measurements were performed on the OSIRIS spectrometer at the ISIS Neutron and Muon Source; the spectrometer configuration and performance are described in Ref.~\cite{telling2005spectroscopic}. The sample was mounted in an aluminum holder. X-ray Laue diffraction was used to assist the initial sample alignment, and the scattering intensity discussed here was measured in the $(H,K,0)$ plane. The INS data discussed here were taken at $T=2$~K, i.e., above the zero-field ordering anomalies, in order to probe the low-energy response outside the ordered phase.

	\begin{table*}[t]
		\caption{\label{tab:cef_levels}Calculated CEF doublets of Er$^{3+}$ in ErBr$_3$. Listed are the excitation energies relative to the ground state and the dominant $|m_J\rangle$ components of a representative partner of each Kramers doublet, obtained by choosing a convenient basis within the degenerate subspace of the McPhase output. The other partner is obtained by time reversal.}
		\begin{ruledtabular}
			\begin{tabular}{@{\hspace{2.5em}}ccc@{\hspace{2.5em}}}
				Doublet & Energy (meV) & Dominant $|m_J\rangle$ components \\
				\hline
				1 (GS) & 0.000 &
				$0.874\ket{15/2}+0.393\ket{9/2}+0.282\ket{3/2}-0.032\ket{-3/2}+0.001\ket{-9/2}$ \\
				2 & 0.480 &
				$0.408\ket{13/2}+0.503\ket{7/2}+0.654\ket{1/2}-0.351\ket{-5/2}+0.174\ket{-11/2}$ \\
				3 & 1.334 &
				$0.484\ket{13/2}+0.303\ket{7/2}-0.056\ket{1/2}+0.640\ket{-5/2}-0.511\ket{-11/2}$ \\
				4 & 1.368 &
				$0.422\ket{15/2}-0.329\ket{9/2}-0.839\ket{3/2}+0.093\ket{-3/2}-0.003\ket{-9/2}$ \\
				5 & 1.876 &
				$0.559\ket{13/2}+0.134\ket{7/2}-0.690\ket{1/2}-0.271\ket{-5/2}+0.346\ket{-11/2}$ \\
				6 & 4.839 &
				$0.461\ket{13/2}-0.685\ket{7/2}+0.167\ket{1/2}-0.342\ket{-5/2}-0.416\ket{-11/2}$ \\
				7 & 4.881 &
				$0.238\ket{15/2}-0.859\ket{9/2}+0.452\ket{3/2}-0.046\ket{-3/2}+0.002\ket{-9/2}$ \\
				8 & 4.928 &
				$0.271\ket{13/2}-0.409\ket{7/2}+0.256\ket{1/2}+0.527\ket{-5/2}+0.645\ket{-11/2}$ \\
			\end{tabular}
		\end{ruledtabular}
	\end{table*}
	
	\begin{table}[t]
		\caption{\label{tab:cef_params}Stevens parameters allowed by the trigonal ($C_3$) site symmetry of the Er$^{3+}$ ion, generated in the semiquantitative McPhase point-charge calculation for Er$^{3+}$ in ErBr$_3$, based on the local atomic positions and nominal ionic charges, following the $B_l^m$ convention of Ref.~\cite{hutchings1964point}.}
		\begin{center}
			\begin{tabular}{@{\hspace{3em}}cD{.}{.}{-1}@{\hspace{3em}}}
				\toprule
				Parameter & \multicolumn{1}{c}{Value (meV)} \\
				\midrule
				$B_2^0$ & -0.005290 \\
				$B_4^0$ & -0.000094 \\
				$B_4^3$ & -0.002681 \\
				$B_6^0$ & 0.000000 \\
				$B_6^3$ & -0.000002 \\
				$B_6^6$ & 0.000002 \\
				\bottomrule
			\end{tabular}
		\end{center}
	\end{table}

	Crystal-field calculations were carried out using the McPhase package \cite{rotter2004using}. In the present work, the calculation is a point-charge calculation based on the local atomic positions and nominal ionic charges around the Er$^{3+}$ ion, rather than a fit of Stevens parameters to the experimental excitation spectrum. It is used mainly to identify the symmetry class and dominant $|m_J\rangle$ components of the low-energy Kramers doublets, while the absolute excitation energies are treated at a semiquantitative level.
	
	\section{Results}
	
	\subsection{Crystal structure and single-ion anisotropy}
	
	ErBr$_3$ crystallizes in the rhombohedral BiI$_3$-type structure with space group $R\bar{3}$ (No.~148) \cite{Kramer1999PRB}. The reported room-temperature lattice parameters are $a=b=7.045$~\AA\ and $c=19.148$~\AA\ \cite{Brown1968}. Its crystal structure is shown in Fig.~\ref{fig:structure_cef}(a), where only the positions of the Er$^{3+}$ ions are shown for clarity. The Er$^{3+}$ ions occupy the Wyckoff $6c$ positions and form well-defined honeycomb layers in the $ab$ plane, while the Br$^-$ ions occupy the $18f$ positions \cite{Brown1968,kaduk2019pearson}. These honeycomb layers stack along the crystallographic $c$ axis in an ABC sequence, giving an interlayer spacing of about 6.4~\AA. As shown in Fig.~\ref{fig:structure_cef}(b), each individual layer forms a regular honeycomb network with a nearest-neighbor Er--Er distance of about 4.1~\AA. The intralayer Er--Er distance is therefore substantially shorter than the crystallographic repeat length along $c$, while the magnetic lattice itself remains layered. ErBr$_3$ thus provides a quasi-two-dimensional honeycomb framework. The dominant magnetic correlations are therefore naturally expected to develop primarily within the $ab$ planes.
	
	The magnetization curves measured at 1.8~K are shown in Fig.~\ref{fig:structure_cef}(c). For fields applied within the honeycomb plane ($H \parallel ab$), the magnetization rises much more rapidly and exceeds $6~\mu_{\mathrm B}$/Er$^{3+}$ at 8.9~T, whereas the response for $H \parallel c$ remains substantially smaller over the entire field range. The $ab$ plane is therefore the magnetic easy direction. This pronounced anisotropy indicates that the magnetic response of ErBr$_3$ is governed by a strongly anisotropic single-ion ground state. It also identifies $H \parallel ab$ as the most relevant tuning direction for the thermodynamic measurements discussed below. 
	
	As shown in Fig.~\ref{fig:structure_cef}(d), each Er$^{3+}$ ion is coordinated by six Br$^-$ ions and forms an ErBr$_6$ octahedron. The corresponding local crystal field has trigonal symmetry and provides the microscopic origin of the single-ion anisotropy discussed below. The Er$^{3+}$ ion has a $4f^{11}$ electronic configuration and total angular momentum $J=15/2$, giving a $2J+1=16$ multiplet. Under the CEF, this multiplet is split into eight Kramers doublets, as illustrated in Fig.~\ref{fig:structure_cef}(e). The low-energy thermodynamic and spectroscopic responses are therefore expected to be governed primarily by the ground-state Kramers doublet, provided that the temperature remains well below the first excited crystal-field level. In this regime, the ground-doublet matrix elements govern the low-energy response. Crystal-field analysis suggests that the ground-state doublet belongs to the symmetry class that can be written as
	%R2(1)
	\begin{align}\label{eq:psi0}
		\psi_{\pm}= &\alpha\ket{\pm 15/2}\pm\beta\ket{\pm 9/2}+\gamma\ket{\pm 3/2}\notag\\
		&\pm\delta\ket{\mp 3/2}+\epsilon\ket{\mp 9/2}\pm\zeta\ket{\mp 15/2}.
	\end{align}
	The two partners are related by time reversal, which fixes the coefficients of the second partner through complex conjugation and the standard phase factors in the $|J,m_J\rangle$ basis.
	
	The CEF excitation energies and the dominant $|m_J\rangle$ components of the corresponding Kramers doublets obtained from the point-charge calculation are summarized in Table~\ref{tab:cef_levels}, which also makes clear the separation between the two symmetry-allowed $|m_J\rangle$ sector patterns discussed below. The corresponding Stevens parameters generated in the point-charge calculation are listed in Table~\ref{tab:cef_params}.
	These tables represent a structure-based point-charge model rather than a unique fit to the experimental spectrum.
	The ground-state doublet belongs to the symmetry class containing only the $m_J=\pm(15,9,3)/2$ sectors and is dominated in amplitude by the $\ket{\pm 15/2}$, $\ket{\pm 9/2}$, and $\ket{\pm 3/2}$ components, consistent with Eq.~\ref{eq:psi0}.
	Although the calculated excitation energies are treated here only at a semiquantitative level, the calculation identifies the wavefunction character of the ground-state doublet and provides a consistent crystal-field framework for the experimentally observed low-energy excitations.
	
	We next examine the symmetry class of this ground-state Kramers doublet and its consequences for the low-energy neutron response. For Er$^{3+}$ ($J=15/2$) in the trigonal local crystal-field environment of ErBr$_3$, the local threefold rotation admixes only $|m_J\rangle$ components differing by $\Delta m_J = 3n$ ($n$ integer) \cite{Abragam.1970,hutchings1964point}, so the CEF states must transform as one of two symmetry-distinct doublet classes under this rotation. These two classes correspond to different allowed sets of $|m_J\rangle$ components in the CEF wave function.
	In particular, one class contains only the $m_J=\pm(15,9,3)/2$ sectors, while the other contains the $m_J=\pm(13,7,1,-5,-11)/2$ sectors. Since the ladder operators $J^\pm$ change $m_J$ by one unit, symmetry directly determines whether the corresponding off-diagonal matrix element within the ground-state doublet is forbidden or allowed.
	One class is represented by Eq.~\ref{eq:psi0}, whereas the other may be written schematically as
	\begin{align}\label{eq:psi1}
		\psi'_{\pm}= &\alpha'\ket{\pm 13/2}\pm\beta'\ket{\pm 7/2}+\gamma'\ket{\pm 1/2}\notag\\
		&\pm\delta'\ket{\mp 5/2}+\epsilon'\ket{\mp 11/2}.
	\end{align}
	The key distinction is how the $|m_J\rangle$ sectors of the two Kramers partners are related. For the first class, Eq.~\ref{eq:psi0}, both partners occupy exactly the same set of sectors, $m_J=\pm(15,9,3)/2$; since $J^\pm$ changes $m_J$ by one unit, it can never connect one member of this three-spaced set to another, so
	\[
	\langle \psi_{\pm} | J^{\pm} | \psi_{\mp} \rangle = 0 .
	\]
	For the second class, Eq.~\ref{eq:psi1}, the two partners instead occupy two different, mirror-image sets of sectors, $m_J=\pm(13,7,1,-5,-11)/2$, whose nearest members differ by only one unit (e.g., $m_J=13/2$ in $\psi'_+$ and $m_J=11/2$ in $\psi'_-$); a single action of $J^\pm$ can therefore bridge the two partners, so
	\[
	\langle \psi'_{\pm} | J^{\pm} | \psi'_{\mp} \rangle \neq 0 .
	\]
	
	The calculated ground-state wavefunction listed in Table~\ref{tab:cef_levels} is not compatible with the second class of Eq.~\ref{eq:psi1}, since its dominant $|m_J\rangle$ sectors follow the pattern of Eq.~\ref{eq:psi0} rather than that of Eq.~\ref{eq:psi1}.
	The calculated ground-state wavefunction therefore belongs to the first class, Eq.~\ref{eq:psi0}, as indicated schematically in Fig.~\ref{fig:structure_cef}(e); this assignment is further consistent with the experimentally observed low-energy CEF excitations.
	The physical consequence can be illustrated by considering the exchange processes involving $J^{\pm}$ operators on a representative nearest-neighbor bond. For example, an isotropic exchange term may be decomposed as
	\[
	H_{ij}^{\rm iso}
	=
	\mathcal{J}\,\mathbf{J}_i\!\cdot\!\mathbf{J}_j
	=
	\mathcal{J}
	\left(
	J_i^zJ_j^z
	+\frac{1}{2}J_i^+J_j^-
	+\frac{1}{2}J_i^-J_j^+
	\right).
	\]
	The last two terms are the transverse exchange terms mediated by $J^\pm$. After projection onto the ground-state doublet, their amplitudes are controlled by the off-diagonal matrix elements $\langle \psi_\pm | J^\pm | \psi_\mp \rangle$.
	In rare-earth CEF systems, the relevant local magnetic operator is the total angular momentum $\mathbf{J}$ rather than the bare spin $\mathbf{S}$. For the ground-state doublet of ErBr$_3$, the relevant local matrix element is therefore $\langle \psi_\pm | J^{\pm} | \psi_\mp \rangle = 0$.
	After projection onto the ground-state doublet, the lowest-order $J^{\pm}$-mediated exchange terms therefore vanish. Closely analogous wavefunction-dependent selection-rule physics has been discussed in TbScO$_3$ \cite{Zhao2023TbScO3} and DyScO$_3$ \cite{Wu2017DyScO3}; by contrast, the corresponding matrix element is finite in YbAlO$_3$ \cite{Wu2019YbAlO3}. TbScO$_3$ hosts a non-Kramers pseudo-doublet rather than the Kramers doublet considered here, so the analogy concerns the projected matrix-element structure and not an identical finite-field response.
	Within the ideal projected ground-state doublet, the same selection rule also gives vanishing in-plane magnetic-dipole matrix elements, so an in-plane field does not produce a linear Zeeman splitting at this lowest-order level. The observed response to $H\parallel ab$ must therefore involve physics beyond the strict ground-doublet projection, such as field-induced admixture with excited crystal-field states and interaction effects. Accordingly, the sharp low-field anomaly is associated with the suppression of the ordered phase, whereas the broader high-field feature is described below using a Schottky-like phenomenological field scale, without assuming that this scale is identical to the microscopic in-plane $g$ factor of the zero-field ground-state doublet.
	
	\begin{figure*}[t]
		\centering
		\includegraphics[width=2\columnwidth]{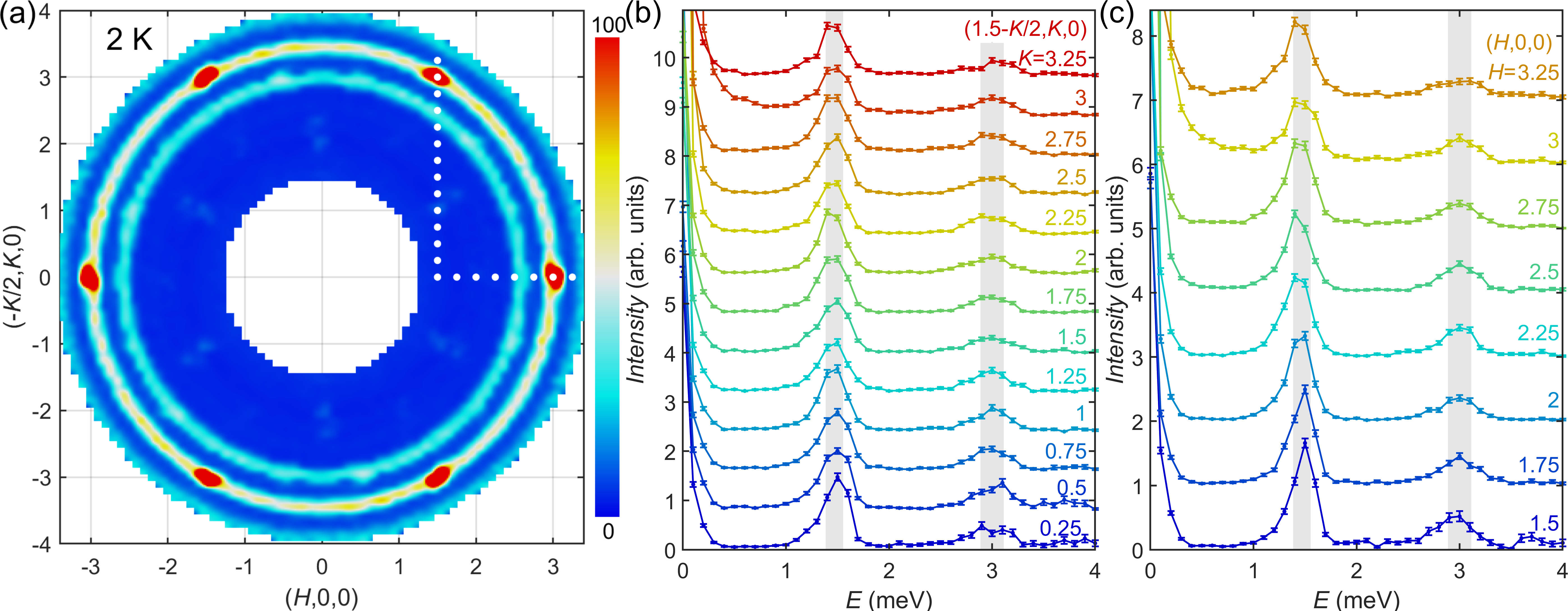}
		\caption{
			(a) Sixfold-symmetrized elastic-channel neutron scattering intensity of ErBr$_3$ measured with an aluminum holder at $T = 2$~K in the $(H,K,0)$ plane. The white dots mark the momentum positions at which the constant-$Q$ energy cuts shown in (b) and (c) were taken. (b),(c) Constant-$Q$ energy cuts measured at representative wave vectors along $(1.5-K/2,K,0)$ and $(H,0,0)$, respectively. Successive curves are vertically offset for clarity. In both momentum directions, the dominant peaks are observed near 1.5 and 3~meV, highlighted by the gray shaded regions, while some cuts also show weak additional intensity at higher energies. The nearly unchanged positions of the dominant peaks for different wave vectors indicate no detectable dispersion within experimental resolution.
		}
		\label{fig:INS_directional_cuts}
	\end{figure*}
	
	The crystal-field excitation spectrum was examined by INS. To probe the dynamical consequences of the selection rule discussed above, we performed INS measurements above the ordering temperature at $T = 2$~K. This temperature lies well above the zero-field ordering anomalies. The measured spectrum therefore primarily reflects the low-energy response of the crystal-field doublet outside the ordered phase. Fig.~\ref{fig:structure_cef}(f) shows the low-energy scattering intensity. The spectrum is dominated by nearly flat crystal-field excitations, and no obvious low-energy magnetic dispersion is resolved. Within the accessible energy and momentum ranges, no well-defined low-energy dispersive magnetic excitations are resolved above background and instrumental broadening. Representative constant-$Q$ energy cuts taken along different momentum directions are shown in Fig.~\ref{fig:INS_directional_cuts}. Their peak positions remain essentially unchanged within experimental resolution, supporting the absence of detectable dispersion. 
	By integrating the intensity over the full measured $Q$ range, we obtain the spectrum shown in Fig.~\ref{fig:structure_cef}(g). A three-peak Lorentzian fit to this spectrum (the third peak accounting for the quasi-elastic intensity near zero energy) gives the dominant crystal-field excitation energies as 1.46 and 2.98~meV, in close agreement with the values of 1.5 and 3.0~meV independently obtained from a fit to inelastic neutron-scattering data in Ref.~\cite{Wessler2022Thesis}.
	For direct comparison, the point-charge calculation gives excited doublets at 0.480~meV, in a group from 1.334 to 1.876~meV, and in a group from 4.839 to 4.928~meV, whereas the dominant observed peaks occur at 1.46 and 2.98~meV. The agreement is therefore not quantitative, and no one-to-one assignment of the measured peaks to the calculated levels is claimed. The point-charge result is used at a semiquantitative level to identify the symmetry class and dominant $|m_J\rangle$ character of the ground-state doublet.
	These dominant nearly nondispersive features are consistent with crystal-field excitations. Although a phonon contribution cannot be excluded solely from the present data, and weaker additional higher-energy structures are also present in parts of the dataset, the weak momentum dependence of the dominant peaks together with the overall crystal-field analysis supports a predominantly CEF origin.
	No coherent mode is resolved within the measured window and experimental sensitivity.
	
	\begin{figure*}[t]
		\centering
		\includegraphics[width=2\columnwidth]{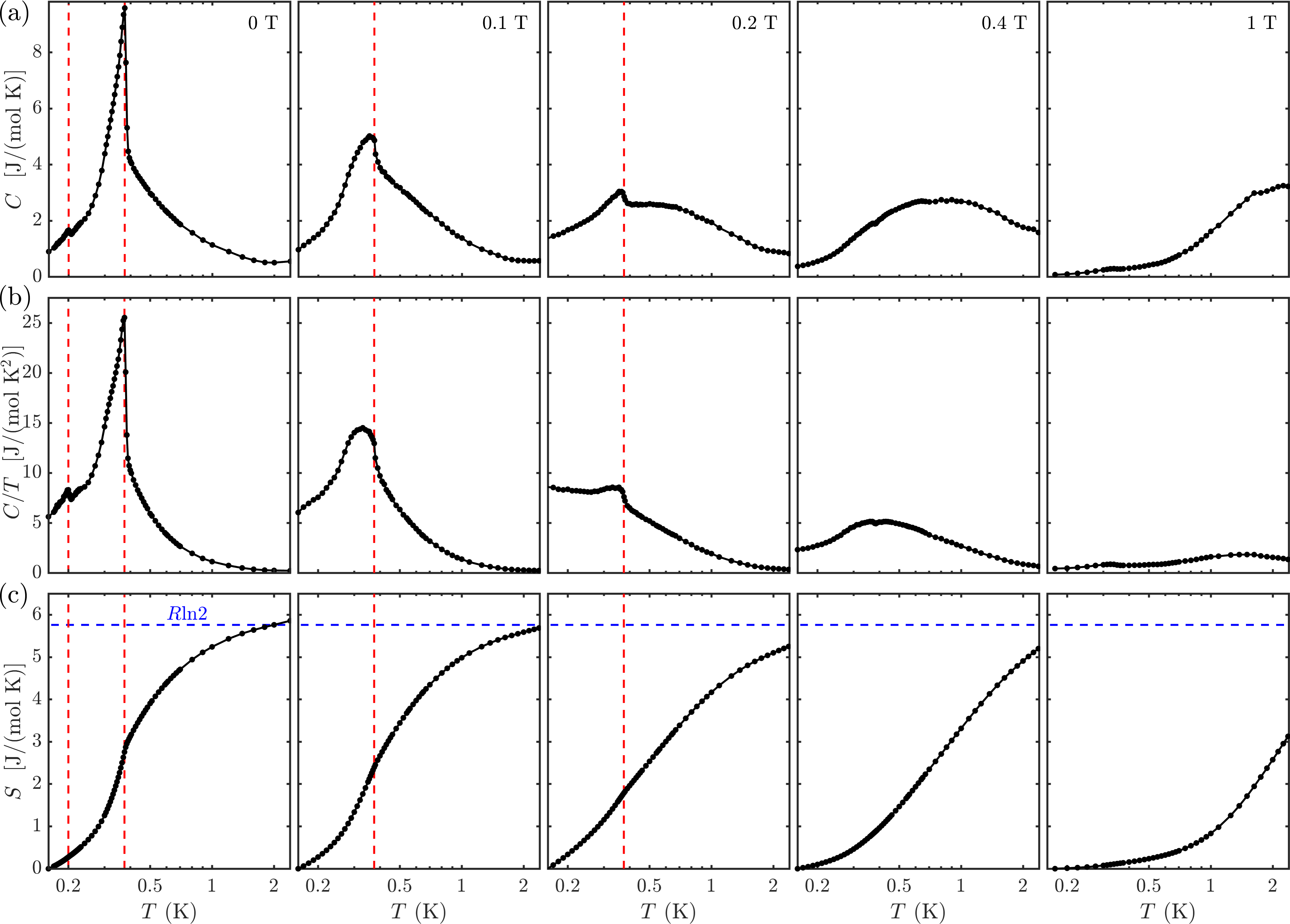}
		\caption{
			(a) Specific heat $C(T)$, (b) $C(T)/T$, and (c) entropy $S(T)$ of ErBr$_3$ at representative magnetic fields applied within the honeycomb plane ($H \parallel ab$).
			Vertical dashed red lines mark the thermodynamic anomalies.
			The horizontal dashed blue line indicates the entropy $R\ln 2$ expected for a Kramers doublet ground state.
		}
		\label{fig:Cp_field}
	\end{figure*}
	
	Previous neutron scattering measurements reported dispersive modes in ErBr$_3$ \cite{Wessler2022}. Those results were analyzed within an MF-RPA framework including dipolar interactions and crystal-field anisotropy \cite{Wessler2022}. The vanishing matrix element $\langle\psi_\pm|J^\pm|\psi_\mp\rangle$ removes only the lowest-order projected $J^{\pm}$-mediated process. Dispersive excitations can remain compatible with this selection rule through higher-order CEF mixing or through interactions such as the dipolar coupling included in that analysis, while weak modes may remain below the present experimental resolution. The present results therefore complement, rather than reinterpret, the excitation spectrum reported in Ref.~\cite{Wessler2022}.
	
	\subsection{Thermodynamics and Phase Diagram}
	
	\begin{figure*}[t]
		\centering
		\includegraphics[width=2\columnwidth]{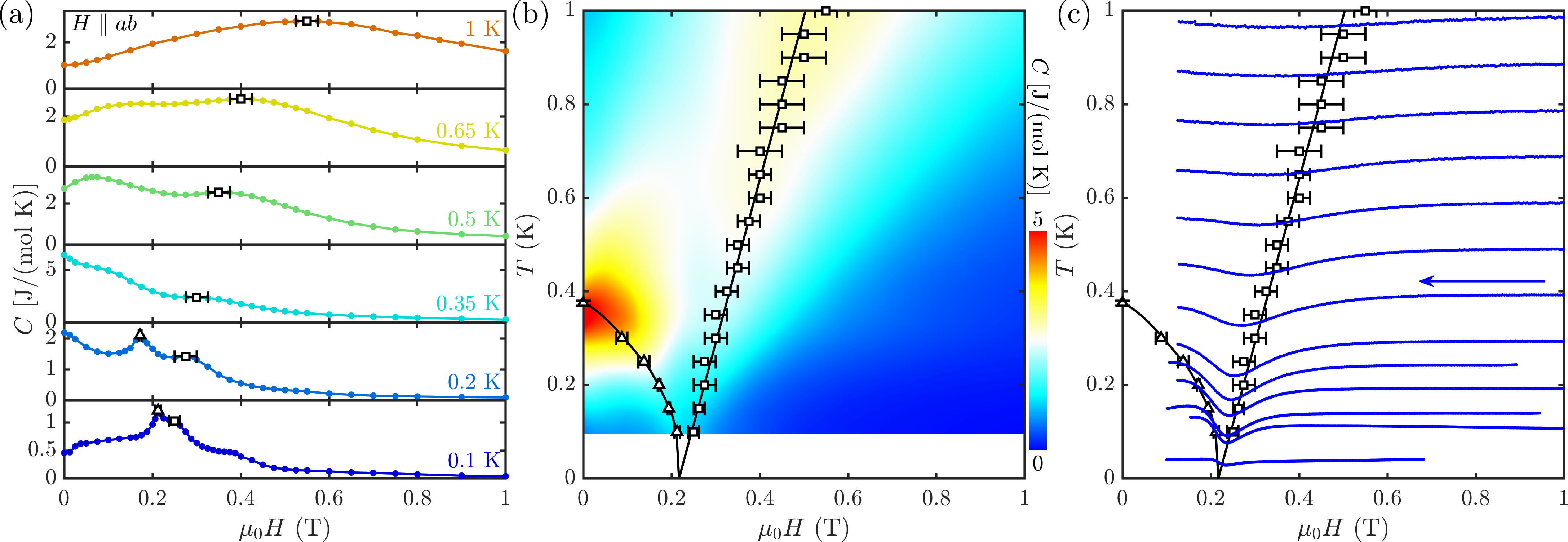}
		\caption{
			Field evolution of the thermodynamic response of ErBr$_3$ for magnetic fields applied within the honeycomb plane ($H \parallel ab$).
			(a) Field dependence of the specific heat measured at selected temperatures. The curves are displayed in separate stacked panels for clarity. Open triangles and open squares mark the phase-boundary and crossover points, respectively. A sharp low-field anomaly associated with the thermodynamic phase boundary is observed together with a broader high-field crossover feature; at 0.35~K, only the crossover feature is resolved as a distinct point, since the phase boundary lies too close to the zero-field transition at 0.375~K to be reliably extracted.
			(b) Field--temperature ($H$--$T$) color map of the specific heat $C$ of ErBr$_3$ for $H \parallel ab$. Open triangles and open squares mark the phase-boundary and crossover points, respectively, as in (a); the triangles are connected by a guide to the eye. The error bars indicate the neighboring temperature or field steps used to bracket the extracted characteristic points. The straight line represents a weighted linear fit to the $T<1$~K crossover points, which deviate from linearity above 1~K, with a slope of 3.49~K/T and a zero-temperature field intercept of 0.22~T.
			(c) Magnetocaloric-effect measurements of ErBr$_3$ during decreasing-field sweeps, shown as blue curves. The markers and guide line from (b) are overlaid for comparison. The strongest response occurs at low temperature near the phase boundary.
		}
		\label{fig:phase_diagram}
	\end{figure*}
	
	The temperature dependence of the specific heat and entropy of ErBr$_3$ for magnetic fields applied within the honeycomb plane ($H \parallel ab$) is shown in Fig.~\ref{fig:Cp_field}. In zero field, two sharp anomalies are resolved at $0.375\pm0.005$~K and $0.200\pm0.002$~K, reflecting the temperature-step resolution near each peak, appearing as a larger peak and a smaller peak, respectively. These two features indicate successive thermodynamic phase transitions at zero field and provide the reference for tracking the field evolution of the thermodynamic response. The lower-temperature anomaly is rapidly suppressed once a finite field is applied, whereas the phase boundary originating from the 0.375-K zero-field anomaly can be followed to about 0.2~T before it is also suppressed. Above about 0.2~T, a broad peak becomes clearly visible in the specific-heat data. Unlike the sharp low-field anomalies, this broad feature does not show the characteristics of a well-defined thermodynamic phase transition. In the following, we therefore treat it as a crossover rather than a distinct phase transition. The entropy curves derived from integrating $C/T$ approach $R\ln 2$ by about 2~K, supporting the picture of a well-isolated Kramers doublet ground state. $S$ is set to zero at 0.16~K, providing a lower bound on the absolute entropy.
	
	In this low-temperature range, the lattice contribution is expected to remain smooth, small, and essentially field independent. At $T=2.4$~K, the highest temperature shown in Fig.~\ref{fig:Cp_field}(b), $C/T\approx0.230$~J/(mol~K$^2$); treating this value as a conservative upper bound on the phonon contribution and extrapolating with the Debye $T^3$ law down to the ordering temperatures gives an estimated phonon entropy of only about $0.18$~J/(mol~K), i.e., roughly 3\% of $R\ln 2$. The phonon contribution at lower temperatures should therefore be negligible for all measured fields. Likewise, the first excited CEF levels lie far above the temperatures of the thermodynamic anomalies, so their Schottky contribution is not expected to generate such narrow low-temperature features. The pronounced low-temperature anomalies and their strong field evolution therefore cannot be attributed primarily to phonons, but instead arise mainly from the low-energy magnetic degrees of freedom within the ground-state doublet.
	
	This is fully consistent with the single-ion crystal-field analysis discussed above. At higher fields, the entropy saturation shifts to higher temperature, indicating that the relevant low-energy scale increases under field. Taken together, these results show that the field response of ErBr$_3$ cannot be reduced to a simple suppression of a single anomaly. Instead, the two zero-field features evolve in clearly different ways under field.
	
	The existence of two sharp anomalies indicates that the zero-field ordered regime is more complex than a single conventional transition. At present, our thermodynamic data establish that both features are genuine low-temperature anomalies, but do not uniquely determine their microscopic origin. One possibility is that they correspond to two successive ordering transitions. Another is that the higher-temperature anomaly marks the onset of long-range magnetic order, whereas the lower-temperature feature reflects a secondary reorganization within the ordered state, such as spin reorientation or an additional effect of weak dipolar interactions. This scenario is broadly consistent with the two- and three-dimensional magnetic ordering reported by powder neutron diffraction in ErX$_3$ compounds, including ErBr$_3$ itself \cite{Kramer1999PRB}. Distinguishing between these scenarios will require further microscopic probes, for example low-temperature neutron diffraction or spectroscopy across both anomalies.
	
	The field dependence of the specific heat at selected temperatures is shown in Fig.~\ref{fig:phase_diagram}(a). The zero-field transition temperature of 0.375~K was determined from the temperature-sweep data, with a vertical error bar reflecting the neighboring temperature steps. The field-sweep data track how the low-field anomaly evolves under applied field.
	Upon cooling from the zero-field transition at 0.375~K, the low-field anomaly moves to finite fields and weakens rapidly, whereas the broader high-field feature remains visible over a wider temperature range. 
	This contrast indicates that the low-field anomaly and the broader high-field feature have different thermodynamic origins rather than reflecting a simple splitting of a single anomaly. For temperatures below this scale, two anomalies can be resolved in the field-sweep data. The low-field anomaly defines the thermodynamic phase boundary, while the broader high-field feature is taken as a crossover. 
	
	The zero-field endpoint at 0.375~K was determined from the temperature-sweep data. The remaining phase-boundary and crossover points were extracted from characteristic features in the field-sweep specific-heat curves. Most points were determined from local maxima; only the 0.10- and 0.35-K shoulder-like high-field features on locally monotonic curves were assigned using the flattest point within the shoulder region. The error bars indicate the neighboring temperature or field steps used to bracket the extracted characteristic points and therefore represent experimental resolution intervals rather than statistical standard deviations.
	
	The positions of these two types of features are collected in the $H$--$T$ color map shown in Fig.~\ref{fig:phase_diagram}(b). The triangular symbols denote the phase boundary, and the square symbols denote the crossover line. The low-temperature crossover points form an approximately linear relation in the $H$--$T$ plane.
	The low-temperature feature is, however, lower than the ideal Schottky peak, and the finite intercept is inconsistent with the simple relation $\Delta=g\mu_{\rm B}B$ for an isolated zero-field doublet. In addition, the point-charge ground-state doublet has vanishing projected in-plane magnetic-dipole matrix elements. The crossover slope is therefore treated as an empirical measure of the field-dependent thermodynamic energy scale. Interactions, weak dispersion, field-induced admixture with excited CEF states, and collective effects may all contribute to the observed field-dependent scale.
	The fact that the thermodynamic phase boundary is confined to much lower fields further suggests that the broad feature reflects a field-driven crossover beyond the low-field ordered regime, consistent with approaching a polarized regime. A rough Curie--Weiss estimate ($\theta_{\rm CW}\approx-8$~K) still exceeds the crossover-slope energy scale, consistent with exchange interactions dominating near the crossover onset.
	
	To verify the thermodynamic nature of the phase boundary, we also performed magnetocaloric-effect (MCE) measurements during continuous decreasing-field sweeps with $H \parallel ab$, as shown in Fig.~\ref{fig:phase_diagram}(c). The phase boundary and crossover trajectory extracted from the specific-heat data are overlaid for comparison. At the lowest temperatures, the MCE curves show the sharpest extrema and the largest slope changes near the phase boundary, indicating particularly strong field-induced entropy changes in this regime. Their positions are consistent with those extracted from the temperature- and field-sweep specific-heat measurements, supporting the thermodynamic nature of the phase boundary. At higher temperatures, the field-sweep curves become weaker and smoother overall, consistent with a reduced MCE response. The strongest MCE response is observed at low temperature near the phase boundary, pointing to potential relevance for sub-Kelvin magnetic refrigeration. Taken together, the specific-heat, entropy, and MCE measurements show that, for $H \parallel ab$, ErBr$_3$ exhibits one thermodynamic phase boundary together with a broader crossover line. The MCE results therefore provide an independent confirmation of the phase line identified from the specific-heat measurements. 
	
	\section{Discussion and Conclusion}
	
	The combined results connect the ground-state doublet symmetry with the thermodynamic and neutron responses of ErBr$_3$. This connection can be followed consistently from the crystal structure and single-ion anisotropy to the thermodynamic phase diagram and neutron response. In ErBr$_3$, the ground-state doublet belongs to a representation with a vanishing off-diagonal matrix element of $J^\pm$. This assignment is also consistent with the local structural environment. The actual ErBr$_6$ octahedron is embedded in a layered lattice and is weakly trigonally distorted from an ideal octahedral environment, with the local triangular faces slightly compressed along the trigonal axis (Fig.~\ref{fig:structure_cef}(d)). Such a distortion provides a natural structural mechanism for selecting between the two symmetry-allowed Kramers-doublet classes discussed above. We verified this ground-state trend over the range of trigonal distortions relevant to ErBr$_3$: compression favors Eq.~\ref{eq:psi0}, elongation favors Eq.~\ref{eq:psi1}. Although the present data do not determine the crystal-field parameters uniquely enough to exclude the alternative class from structure alone, the local structural environment, together with the absence of a clear low-energy dispersive mode above the ordering temperature, supports the $\pm(15,9,3)/2$ assignment. No well-defined low-energy dispersive magnetic mode is resolved within the experimental sensitivity; this observation does not by itself establish the absence of magnetic excitations. The present results complement, rather than reinterpret, the dispersive modes reported inside the ordered phase \cite{Wessler2022}.
	
	This selection rule suppresses the lowest-order exchange processes mediated by $J^{\pm}$ operators within the ground-state doublet. Separately, the thermodynamic measurements reveal a phase boundary together with a broader crossover whose characteristic temperature increases approximately linearly with field. We report this field dependence through its measured slope and intercept without identifying it with a microscopic in-plane $g$ factor of the zero-field projected doublet.
	Small discrepancies in the measured zero-field specific heat between Figs.~\ref{fig:Cp_field} and \ref{fig:phase_diagram}, larger near the two anomalies, reflect session-to-session variations and do not affect the conclusions drawn here.
	
	Related strongly anisotropic rare-earth ground states have been reported in TbScO$_3$ and DyOCl \cite{Zhao2023TbScO3,Tian2021DyOCl}.
	More broadly, they show that the crystal-field symmetry of the ground-state doublet can strongly influence both the excitation spectrum and the field-induced phase evolution in spin--orbit--entangled rare-earth magnets. Field-tuned thermodynamic behavior and quantum criticality have also been investigated in honeycomb-lattice magnets including YbCl$_3$ and K$_4$MnMo$_4$O$_{15}$ \cite{Matsumoto2024YbCl3,Khatua2025K4MnMo4O15}, motivating further studies of the associated critical behavior in ErBr$_3$.
	ErBr$_3$ thus provides a useful platform for relating single-ion symmetry constraints to collective behavior in a two-dimensional honeycomb magnet.
	
	\section*{Acknowledgments}
	
	This work was supported by the National Key Research and Development Program of China (Grant Nos.~2021YFA1400400 and 2025YFA1411503), 
	the National Natural Science Foundation of China (Grant No.~12374146), 
	the Guangdong Basic and Applied Basic Research Foundation (Grant No.~2024B1515120045), 
	the Guangdong Provincial Quantum Science Strategic Initiative (Grant No.~GDZX2401006), 
	the Shenzhen Fundamental Research Program (Grant No.~JCYJ20220818100405013), 
	the Open Research Fund of State Key Laboratory of Quantum Functional Materials (Grant No.~QFM2025KF007), 
	and the Guangdong Provincial Key Laboratory of Advanced Thermoelectric Materials and Device Physics (Grant No.~2024B1212010001). 
	
	Experiments at the ISIS Neutron and Muon Source were supported by beamtime allocation RB2510565 from the Science and Technology Facilities Council. A.A.P. acknowledges the support of the Scientific User Facilities Division, Office of Basic Energy Sciences, U.S. Department of Energy. We acknowledge the support of ISIS technical staff, thank George~D.~A.~Wood for valuable comments and assistance with the neutron-scattering experiments, and thank Han~Ge for useful discussions and assistance with physical-property measurements.
	
	\bibliography{ErBr3_2610}
	
\end{document}